%%%%%%%%%%%%%%%%%%%%%%%%%%%%%%%%%%%%%%%%%%%%%%%%%%%%%%%%%%%%%%%%%%%%
%  TeX Definitions                                                 %
%%%%%%%%%%%%%%%%%%%%%%%%%%%%%%%%%%%%%%%%%%%%%%%%%%%%%%%%%%%%%%%%%%%%

\newif\iffigs\figstrue
% Uncomment the next line if you do not want the figures
%\figsfalse

%
% the following is to use blackboard bold fonts --
\let\useblackboard=\iftrue
%
% activate this if you don't have them.
%\let\useblackboard=\iffalse
%
% You might also need to remove this line.
\newfam\black

\input harvmac.tex

\iffigs
  \input epsf
\else
  \message{No figures will be included.  See TeX file for more
information.}
\fi

\def\Title#1#2{\rightline{#1}
\ifx\answ\bigans\nopagenumbers\pageno0\vskip1in%
\baselineskip 15pt plus 1pt minus 1pt
\else%\special{papersize=11in,8.5in}%
\def\listrefs{\footatend\vskip 1in\immediate\closeout\rfile\writestoppt
\baselineskip=14pt\centerline{{\bf References}}\bigskip{\frenchspacing%
\parindent=20pt\escapechar=` \input
refs.tmp\vfill\eject}\nonfrenchspacing}
\pageno1\vskip.8in\fi \centerline{\titlefont #2}\vskip .5in}

\ifx\answ\bigans\def\tcbreak#1{}\else\def\tcbreak#1{\cr&{#1}}\fi
\useblackboard
\message{If you do not have msbm (blackboard bold) fonts,}
\message{change the option at the top of the tex file.}
\font\blackboard=msbm10 %scaled \magstep1
\font\blackboards=msbm7
\font\blackboardss=msbm5
%\newfam\black
\textfont\black=\blackboard
\scriptfont\black=\blackboards
\scriptscriptfont\black=\blackboardss
\def\Bbb#1{{\fam\black\relax#1}}
\else
\def\Bbb#1{{\bf #1}}
\fi
% *************************************
%
\def\yboxit#1#2{\vbox{\hrule height #1 \hbox{\vrule width #1
\vbox{#2}\vrule width #1 }\hrule height #1 }}
\def\fillbox#1{\hbox to #1{\vbox to #1{\vfil}\hfil}}
\def\ybox{{\lower 1.3pt \yboxit{0.4pt}{\fillbox{8pt}}\hskip-0.2pt}}

\def\comments#1{}

\def\half{{1\over 2}}

\def\II{\relax{I\kern-.07em I}}

\def\IZ{\relax\ifmmode\mathchoice
{\hbox{\cmss Z\kern-.4em Z}}{\hbox{\cmss Z\kern-.4em Z}}
{\lower.9pt\hbox{\cmsss Z\kern-.4em Z}}
{\lower1.2pt\hbox{\cmsss Z\kern-.4em Z}}\else{\cmss Z\kern-.4em
Z}\fi}
\def\IB{\relax{\rm I\kern-.18em B}}

\def\ID{\relax{\rm I\kern-.18em D}}
\def\IE{\relax{\rm I\kern-.18em E}}
\def\IF{\relax{\rm I\kern-.18em F}}
\def\IG{\relax\hbox{$\inbar\kern-.3em{\rm G}$}}
\def\IGa{\relax\hbox{${\rm I}\kern-.18em\Gamma$}}
\def\IH{\relax{\rm I\kern-.18em H}}
\def\II{\relax{\rm I\kern-.18em I}}
\def\IK{\relax{\rm I\kern-.18em K}}
\def\IP{\relax{\rm I\kern-.18em P}}
%\def\IX{\relax{\rm X\kern-.01em X}}
%this doesn't work

\useblackboard
\def\IZ{\relax\Bbb{Z}}
\fi

\font\cmss=cmss10 \font\cmsss=cmss10 at 7pt
\def\IR{\relax{\rm I\kern-.18em R}}

\def\BR{\IR}

\def\BR{\IR}

\def\tilde{\widetilde}

%%%%%%%%%%%%%%%%%%%%%%%%%%%%%%%%%%%%%%%%%%%%%%%%%%%%%%%%%%%%%%%%%%%%%%%%%%%%
%                    F I G U R E S                                         %
%%%%%%%%%%%%%%%%%%%%%%%%%%%%%%%%%%%%%%%%%%%%%%%%%%%%%%%%%%%%%%%%%%%%%%%%%%%%
%\figsfalse

\iffigs
  \input epsf
\else
  \message{No figures will be included.  See TeX file for more
information.}
\fi

%%% \iffigs
%%% \midinsert
%%% $$\vbox{\centerline{\epsfxsize=4in\epsfbox{figa.eps}}
%%% \centerline{Figure 1. $E_2$ and related theories.}}$$
%%% \endinsert
%%% \fi

%%%%%%%%%%%%%%%%%%%%%%%%%%%%%%%%%%%%%%%%%%%%%%%%%%%%%%%%%%%%%%%%%%%%%%%%%%%%
%                    Definitions from LaTeX                                %
%%%%%%%%%%%%%%%%%%%%%%%%%%%%%%%%%%%%%%%%%%%%%%%%%%%%%%%%%%%%%%%%%%%%%%%%%%%%

%%%
%%% All those have problems with Font \rm
%%%

\def\lim{{lim}}

%%%%%%%%%%%%%%%%%%%%%%%%%%%%%%%%%%%%%%%%%%%%%%%%%%%%%%%%%%%%%%%%%%%%%%%%%%%%
%                    My definitions                                        %
%%%%%%%%%%%%%%%%%%%%%%%%%%%%%%%%%%%%%%%%%%%%%%%%%%%%%%%%%%%%%%%%%%%%%%%%%%%%
\input epsf

                   % N=? SUSY
                             % [
                             % ]

                              % wedge product

                              % Wilson lines

                              % inverse
                           % O(x)

\def\MR#1{{{\BR}^{#1}}}               % Real numbers
               % Complex numbers

%%% \def\MR#1{{{\bf R}^{#1}}}               % Real numbers
%%% \def\MC#1{{{\bf C}^{#1}}}               % Complex numbers
\def\MR#1{{{\BR}^{#1}}}               % Real numbers
               % Complex numbers
               % Circle, sphere,...
               % disk, ball,...
               % Torus
              % CP
               % Ruled surface F_n

             % Patch
                    % line-bundle
              % derivative

                 % Left large bracket
                % Right large bracket
              % SL(*,Z)

                             % identity matrix

      % commutator
               % anti-commutator

           % expectation value
    % expectation value of trace

      % trace
            % trace
            % Trace
            % trace in a rep
            % Trace in a rep

                      % representation
                  % Imaginary
                  % Imaginary

%\def\widebar#1{{\bar{#1}}}                    % Wide bar
                    % Wide bar
                 % Pauli matrix

                      % correction O()
                     % Normal bundle

                      % Hodge star
                         % sign
\def\hepth#1{{\it hep-th/{#1}}}

\def\frac#1#2{{{{#1}}\over {{#2}}}}           % {} over {}

\def\e8{E_8 \times E_8}                       %E_8 times E_8
\def\HALFSONE{{\bf S}^{1}/{\bf Z}_2}    %S^1 over Z_2
%%%%%%%%%%%%%%%%%%%%%%%%%%%%%%%%%%%%%%%%%%%%%%%%%%%%%%%%%%%%%%%%%%%%%%%%%%%%
%                    Greek                                                 %
%%%%%%%%%%%%%%%%%%%%%%%%%%%%%%%%%%%%%%%%%%%%%%%%%%%%%%%%%%%%%%%%%%%%%%%%%%%%

% \def\a{{\alpha}}              % Appears above

\def\lam{{\lambda}}

%%% \def\Dsh{{D\!\!\!\slash}}     % D slash

%%%%%%%%%%%%%%%%%%%%%%%%%%%%%%%%%%%%%%%%%%%%%%%%%%%%%%%%%%%%%%%%%%%%%%%%%%%%
%     Special Purpose  Definitions                                         %
%%%%%%%%%%%%%%%%%%%%%%%%%%%%%%%%%%%%%%%%%%%%%%%%%%%%%%%%%%%%%%%%%%%%%%%%%%%%
                            % 2\times 2  J

                                % Wave function propagator

%%% ------------------------- CUT HERE ---------------------------------%

%%%%%%%%%%%%%%%%%%%%%%%%%%%%%%%%%%%%%%%%%%%%%%%%%%%%%%%%%%%%%%%%%%%%%%%%%%%%
%                    TITLE PAGE                                            %
%%%%%%%%%%%%%%%%%%%%%%%%%%%%%%%%%%%%%%%%%%%%%%%%%%%%%%%%%%%%%%%%%%%%%%%%%%%%

% \draftmode

%
\Title{ \vbox{\baselineskip12pt\hbox{hep-th/9801034, PUPT-1756}}}
{\vbox{
\centerline{A Matrix Model for Heterotic Spin(32)/$Z_2$}
\centerline{and Type I String Theory.}}}
\centerline{Morten Krogh}
\smallskip
\smallskip
\centerline{Department of Physics, Jadwin Hall}
\centerline{Princeton University}
\centerline{Princeton, NJ 08544, USA}
\centerline{\tt krogh@princeton.edu}
%%%
\bigskip
\bigskip
\noindent
We consider Heterotic string theories in the DLCQ. We 
derive that the matrix model of the Spin(32)/$Z_2$ Heterotic theory 
is the theory living on N D-strings in type I wound on a circle with 
no Spin(32)/$Z_2$ Wilson line on the circle. This is an O(N) gauge theory.
We rederive the matrix model for the $\e8$ Heterotic string theory, explicitly 
taking care of the Wilson line around the lightlike circle. 
The result is the same theory as for Spin(32)/$Z_2$ except that 
now there is a Wilson line on the circle. We also see that the 
integer N labeling the sector of the O(N) matrix model is not just the 
momentum around the lightlike circle, but a shifted momentum 
depending on the Wilson line. We discuss the aspect of level matching,
GSO projections and why, from the point of view of matrix theory
 the $\e8$ theory, and not the Spin(32)/$Z_2$, 
 develops an 11'th dimension for strong coupling. Furthermore 
a matrix theory for type I is derived. This is again the O(N) 
theory living on the D-strings of type I. For small type I 
coupling the system is 0+1 dimensional quantum mechanics.

\Date{January, 1998}

%%% ------------------------- CUT HERE ---------------------------------%

%%%%%%%%%%%%%%%%%%%%%%%%%%%%%%%%%%%%%%%%%%%%%%%%%%%%%%%%%%%%%%%%%%%%
%  B I B L I O G R A P H Y                                         %
%%%%%%%%%%%%%%%%%%%%%%%%%%%%%%%%%%%%%%%%%%%%%%%%%%%%%%%%%%%%%%%%%%%%

\lref\PolWit{Joseph Polchinski and Edward Witten,
  {\it ``Evidence for Heterotic- Type I String Duality,''}
    \hepth{9510169}}

\lref\Connes{Alain Connes, Michael R. Douglas and Albert Schwarz,
  {``Noncommutative Geometry and Matrix Theory: Compactification on 
     Tori,''} \hepth{9711162}}

\lref\DougHull{Michael R. Douglas and Chris Hull,
  {\it ``D-branes and the Noncommutative Torus,''}
  \hepth{9711165}}

\lref\DVV{R.Dijkgraaf, E. Verlinde and H. Verlinde,
  {\it ``Matrix String Theory,''}
  \hepth{9703030}}

\lref\Pioline{N.A. Obers, B. Pioline and E. Rabinovici, 
  {\it ``M-theory and U-duality on $T^d$ with Gauge Backgrounds,''} 
   \hepth{9712084}}

\lref\Narain{K. Narain, M. Sarmadi and E. Witten,
  Nucl. Phys. B279 (1987) 369.}

\lref\Ginsparg{P Ginsparg,
  Phys. Rev. D35 (1987) 648.}

\lref\Sen{Ashoke Sen,
  {\it ``D0 Branes on $T^n$ and Matrix Theory,''} \hepth{9709220}}

\lref\MotlBanks{Tom Banks and Lubos Motl,
  {\it ``Heterotic Strings from Matrices,''}
  \hepth{9703218}}

\lref\BFSS{T. Banks, W. Fischler, S.H. Shenker and L. Susskind,
  {\it ``M Theory As A Matrix Model: A Conjecture,''} \hepth{9610043}}

\lref\Horava{Petr Horava,
  {\it ``Matrix Theory and Heterotic Strings on Tori,''} \hepth{9705055}}

\lref\Lowe{David A. Lowe,
  {\it ``Heterotic Matrix String Theory,''}
  \hepth{9704041}}

\lref\Kleb{Ulf Danielsson, Gabriele Ferretti and Igor R. Klebanov,
  {\it ``Creation of Fundamental strings by crossing D-branes,''} 
    \hepth{9705084}}

\lref\Ulf{Ulf Danielsson and Gabriele Ferretti,
  {\it ``Creation of strings in D-particle Quantum Mechanics,''}
     \hepth{9709171}}

\lref\Kim{Nakwoo Kim and Soo-Jong Rey,
  {\it ``M(atrix) Theory on an Orbifold and Twisted Membrane,''}
    \hepth{9701139}}

\lref\Govin{Suresh Govindarajan,
  {\it ``Heterotic M(atrix) theory at generic points in Narain moduli space,''}
  \hepth{9707164}}

\lref\Seiwhy{N. Seiberg,
  {\it ``Why is the Matrix Model Correct?,''} \hepth{9710009}}

\lref\Eva{Shamit Kachru and Eva Silverstein,
  {\it ``On Gauge Bosons in the Matrix Model Approach to M Theory,''}
  \hepth{9612162}}

\lref\Lubos{Lubos Motl and Leonard Susskind,
  {\it ``Finite N Heterotic Matrix Models and Discrete Light Cone
   Quantization,''}
  \hepth{9708083}}

\lref\Motlscrew{Lubos Motl,
  {\it ``Proposals on nonperturbative superstring interactions,''}
  \hepth{9701025}}

\lref\BS{Tom Banks and Nathan Seiberg,
  {\it ``Strings from Matrices,''}
  \hepth{9702187}}

\lref\Rey{Soo-Jong Rey,
  {\it ``Heterotic M(atrix) Strings and Their Interactions,''}
  \hepth{9704158}}

\lref\Sav{Soo-Jong Rey and Savdeep Sethi,
  { to appear.}}

\lref\Banksrev{Tom Banks,
  {\it ``Matrix Theory,''}
  \hepth{9710231}}

%%% ------------------------- CUT HERE ---------------------------------%

% ===================================================================== %
% Introduction
% ===================================================================== %
\newsec{Introduction}

A matrix model for the $E_8 \times E_8$ Heterotic string theory has been 
developed over the past year \refs{\Eva, \Kim, \MotlBanks, \Lowe, \Rey, \Horava,
 \Govin, \Lubos}. The model can be derived by following Seiberg's and 
 Sen's prescription \refs{\Sen , \Seiwhy } for M-theory on $\HALFSONE$. The 
result is that Heterotic $E_8 \times E_8$ in the sector N of DLCQ is 
described by the decoupled theory of N D-strings in type I wound on a 
circle with a Wilson line on the circle breaking $SO(32)$ to $SO(16) \times 
SO(16)$. It is important to note that this matrix theory is a description 
of the $\e8$ Heterotic string with a Wilson line on the lightlike circle, 
which breaks $\e8$ down to $SO(16) \times SO(16)$. Later we will see that 
the integer N is not the pure momentum around the lightlike circle but a 
shifted momentum depending on winding and $\e8$ charges.

In this paper we will follow Seiberg's prescription for the $Spin(32)/{Z_2}$
Heterotic string theory in the DLCQ and derive a matrix description of it. 
This time we do not take a Wilson line on the lightlike circle.  
The resulting matrix model is again the theory of N D-strings in type I 
wound on a circle but this time without a Wilson line. Since type I and 
Heterotic $SO(32)$ are dual in ten dimensions this also gives a matrix 
model for the type I string theory.

The organization of the paper is as follows. In section 2 we derive the 
matrix model for the $SO(32)$ Heterotic string. In section 3 we rederive 
along similar lines the well known result for the $\e8$ Heterotic string.       
In the process we will get an understanding of the integer N labeling 
the rank of the matrices. In section 4 we will discuss various aspects 
of the two Heterotic matrix models, such as level matching, 
GSO projections and why the $\e8$ theory and not the $SO(32)$ 
develops an extra dimension for strong coupling.
In section 5 we will use the type I - Heterotic $SO(32)$ duality
to derive a matrix model for type I. It turns out that type I
perturbation theory is described by a 0+1 dimensional quantum mechanics
matrix model. We will briefly describe an intuitive picture for 
understanding type I string theory. 

We were informed that a different approach to a matrix model of type I
 and Heterotic Spin(32)/$Z_2$ 
string theory is being taken in a forthcoming paper \Sav.

%%% ------------------------- CUT HERE ---------------------------------%

% ===================================================================== %
% Section (2): The $Spin(32)/ Z_2$ Heterotic string
% ===================================================================== %
\newsec{The Spin(32)/$Z_2$ Heterotic string}

Let us consider the $SO(32)$ Heterotic string with string scale M and 
coupling $\lam$. We compactify it on a lightlike circle of radius R 
with no Wilson line and consider the sector with momentum N. We really 
take this to mean that the theory is compactified on an almost lightlike 
circle as explained in \Seiwhy\ . By a boost and a rescaling of the mass 
scale this takes us to a spatial compactification on a circle of length
$R_s$ and string scale $m_s$. We send $R_s \rightarrow 0$ keeping 
\eqn\fikse{ 
m_s^2 R_s = M^2 R 
} 
The momentum around the circle is $ N \over R_s $. The resulting theory 
in the limit is our answer. We will now apply various well established 
string dualities to obtain a simple description of the answer.

 First we perform a T-duality on the circle. This is simple since there 
is no Wilson line. The momentum number N turns into fundamental string 
winding number N. The T-dual theory has coupling $\lambda^{'}$, string
 scale $m_s^{'}$ and radius $R_s^{'}$ given by
\eqn\dualen{\eqalign{
m_s^{'} &= m_s \cr
R_s^{'} &= {1 \over {m_s^2 R_s}} \cr
\lambda^{'} &= {\lambda \over {m_s R_s}}
}}
Next we employ the ten dimensional type I - SO(32) Heterotic duality
which turn fundamental Heterotic strings into D-strings of type I. 
We thus obtain a type I theory with N D-strings wound on a circle. 
The coupling $\lambda^{''}$, string scale $m_s^{''}$ and radius $R_s^{''}$ 
are 
\eqn\dualto{\eqalign{
m_s^{''} &= {m_s^{'} \over {\sqrt{\lambda^{'}}}} 
= m_s^{3 \over 2} R_s ^{\half} \lambda^{- \half} \cr
R_s^{''} &= R_s^{'} = {1 \over {m_s^2 R_s}} = {1 \over {M^2 R}} \cr
\lambda^{''} &= {1 \over \lambda^{'}} = {{m_s R_s} \over \lambda} 
}}
There is still no Wilson line on the circle. We see that in the limit 
$R_s \rightarrow 0$ with $m_s^2 R_s$ fixed
\eqn\dekobling{\eqalign{
\lambda^{''} & \rightarrow 0 \cr
m_s^{''} & \rightarrow \infty
}}
This is exactly the limit in which the theory on the D-strings decouple 
from the bulk and is described by a SYM theory with $(0,8)$ supersymmetry.
 This theory 
is well known to be an O(N) gauge theory with the following field content

\item{1.} 8 scalars $X^i \;\; i=1,\ldots,8$ in the symmetric part of the 
$N \otimes N$ of O(N) and in the vector representation of the transverse 
spin(8).
\item{2.} 8 rightmoving fermions $\Theta_+$ in the symmetric part of $N 
\otimes N$ of O(N) and in the spinor $8_c$ of spin(8).
\item{3.} The gauge field $A_{\mu}$ in the adjoint of O(N).
\item{4.} 8 leftmoving fermions $\lambda_-$ in the adjoint of O(N) and 
$8_s$ of spin(8).
\item{5.} Leftmoving fermionic fields, $\chi$, in the fundamental of O(N) and 
fundamental of the global SO(32).

The $\chi$ come from 1-9 strings. The rest come from 1-1 strings. The gauge
 coupling is\footnote{$^{1}$}{Pure numbers and factors of $\pi$ are mostly ignored 
in this paper}
\eqn\gaugekobling{
{1 \over {g^2}} = { \lambda^2 \over {M^4 R^2}}
}
The action is standard and can be found in \Lowe.

%%% ------------------------- CUT HERE ---------------------------------%

% ===================================================================== %
% Section (3): Heterotic E_8 
% ===================================================================== %
\newsec{Heterotic $\e8$ String theory}

In this section we will derive the well known result for the matrix 
model of the $\e8$ Heterotic string in the DLCQ. We will follow a 
similar route to the SO(32) case. We take the theory on a lightlike 
circle of radius R with mass scale M and coupling $\lam$. This time 
we put a Wilson line along the lightlike circle. We take the one 
which breaks $\e8$ down to $SO(16) \times SO(16)$. We put momentum
$N \over R$ along the circle. By a boost and rescaling we get to a 
spatial compactification with radius $R_s$ and mass scale $m_s$. Again 
we take $R_s \rightarrow 0$ with 
\eqn\fixeto{m_s^2 R_s = M^2 R.}
The spatial circle now has the Wilson line.

We do not think of the $\e8$ theory as M-theory on $\HALFSONE$ as is usually 
done. The problem with this approach is that with the present knowledge 
it is impossible to treat the Wilson line rigorously. By shrinking the 
spatial circle M-theory on $\HALFSONE$ turns into type IA theory. However 
type IA requires the specification of the positions of the 8-branes. 
Obviously these positions are determined by the Wilson line on the 
shrinking spatial circle, but the exact rules for this are not known yet 
and one is forced to the reasonable guess that the 8-branes are split 
evenly between the 2 orientifold planes.

Instead of regarding $\e8$ Heterotic theory as M-theory on $\HALFSONE$ 
we will use string dualities to obtain the result. First we apply a 
T-duality into the SO(32) Heterotic string on a circle with a Wilson 
line breaking SO(32) to $SO(16) \times SO(16)$. This theory has 
coupling $\lam^{'}$, string scale $m_s^{'}$ and radius $R_s^{'}$ given 
by\refs{\Narain, \Ginsparg }     
\eqn\dualtre{\eqalign{
m_s^{'} &= m_s \cr
R_s^{'} &= {1 \over {2 m_s^2 R_s}} \cr
\lambda^{'} &= {\lambda \over {\sqrt{2} m_s R_s}}
}}
Note the extra factor of 2 compared to usual $R \rightarrow {1 \over {m_s^2 R}}$
T-duality.

We expect a T-duality to take momentum around the circle into fundamental 
string winding. However that is not completely correct in this case. The 
Heterotic $\e8$ theory on a circle is described by a lattice 
of signature (17,1). States in the theory are labelled by points in the 
lattice. A point in the lattice can be specified by the momentum around 
the circle, $N$, winding, $m$, and a lattice vector,$P$, in the weight lattice
 of $\e8$. Furthermore let A be the vector in $\MR{16}$ specifying the 
Wilson line on the circle. Here $\MR{16}$ is to be thought of as the 
vector space spanned by the $\e8$ lattice. The result of T-duality is that 
the winding number, $\tilde N$, in the SO(32) string theory is
\eqn\skiftetmom{
\tilde{N} = 2 (N-m-A \cdot P)
}
The factor of two is related to the extra factor of two in \dualtre.
$\tilde N$ is always an integer because $A \cdot P$ is either integer or 
half-integer. $A \cdot P$ is integer and $\tilde N$ is even if P is a weight 
such that the state is in a true representation (not spinor) of either 
both SO(16)'s or none of them. Otherwise $A \cdot P$ is half-integer and $\tilde N$ 
is odd. For instance the adjoint of $\e8$ splits under $SO(16) \times 
SO(16)$ into (120,1) + (1,120) +(128,1) + (1,128). The two first terms 
correspond to states with even $\tilde N$ and the last two to states 
with odd $\tilde N$.

Now we again employ Heterotic SO(32)- type I duality to get the type I 
theory with $\tilde N$ D-strings wound on a circle of radius $R_s^{''}$,
string scale, $m_s^{''}$, and coupling, $\lam^{''}$ given by
\eqn\dualfire{\eqalign{
m_s^{''} &= {m_s^{'} \over {\sqrt{\lambda^{'}}}} 
= 2^{1 \over 4} m_s^{3 \over 2} R_s ^{\half} \lambda^{- \half} \cr
R_s^{''} &= R_s^{'} = {1 \over {2 m_s^2 R_s}} = {1 \over {2 M^2 R}} \cr
\lambda^{''} &= {1 \over \lambda^{'}} = {{\sqrt{2} m_s R_s} \over \lambda} 
}}
There is still a Wilson line on the circle breaking the type I SO(32) to 
$SO(16) \times SO(16)$.
The limit $R_s \rightarrow 0$ works as before
\eqn\dekoblingto{\eqalign{
\lambda^{''} & \rightarrow 0 \cr
m_s^{''} & \rightarrow \infty
}}
and the theory on the $\tilde N$ D-strings decouple. We get almost the 
same theory as in section 2. The field content is the same and the gauge 
group is $O(\tilde N)$. The difference is the Wilson line which now means
 that half of the 32 fermions have opposite boundary conditions on going 
around the circle. The fermions, $\chi$, transform in the $(\tilde N,32)$ 
of $O(\tilde N) \times SO(32)$. There is a dynamical $O(\tilde N)$ gauge field,
$A_{\mu}$. The SO(32) holonomy is fixed and it  
multiplies half of the fermions, $\chi$, by -1 on going once around the circle.

In going through these dualities we have recovered the well known result 
for this matrix model. However we now understand better the role of the 
integer $\tilde N$ in $O(\tilde N)$. It is not equal to the momentum 
around the lightlike circle but a generalization of it given by eq.
\skiftetmom. This also explains the observation in \Eva, that $\e8$ gauge 
transformations relate sectors with different values of $\tilde N$. 
In Seiberg's prescription states with negative momentum 
around the small spatial circle decouple. The equivalent criterium for 
decoupling is here $\tilde N <0$. Especially we see that states with 
negative momentum N can have $\tilde N >0$ if m and P are chosen 
properly. This result about the change of interpretation of the integer 
labeling a sector have also been noted by \Pioline\ in the case of 
M-theory on a torus.

%%% ------------------------- CUT HERE ---------------------------------%

% ===================================================================== %
% Section (4) Aspects of the Heterotic theories
% ===================================================================== %
\newsec{Aspects of the Heterotic theories}

In this section we will discuss some aspects of the theories of 
the previous sections.
Let us first discuss level matching. 
Because of S-duality a D-string behaves like a fundamental string. The 
D-strings in question here are wound on a circle, and we are describing 
them in static gauge. This corresponds to a fundamental string wound on 
a circle. Let us review the level matching condition for a wound fundamental 
string. Let the string have winding number, n, and momentum number, m. The 
oscillator number on the rightmoving side is called M and on the leftmoving 
side $\bar M$. These include possible zero point energies. Let the circle have 
radius R and the string mass be $m_s$. Now the right- and leftmoving Virasoro 
generators are
\eqn\vira{\eqalign{
L_0 &= {1 \over 4}({m \over R} - n m_s^2 R)^2 + M m_s^2 \cr
{\bar L_0} &= {1 \over 4}({m \over R} + n m_s^2 R)^2 + {\bar M} m_s^2 
}}
The level matching condition is now $L_0 = {\bar L_0}$. This can also be 
written
\eqn\levelmatch{
M - {\bar M} = mn
}
The energy of such a state is 4$ L_0 $. The situation we are 
interested in is where the winding string (the D-string) has a tension that goes 
to infinity and R is fixed. Here the main contribution to the energy comes from 
the winding. We are interested in the total energy minus the winding energy. This
difference is the 
DLCQ $m^2 \over {2R}$. This energy is easily calculated to be
\eqn\energi{
{\rm Energy - winding \; energy} = {{{\bar M} + M} \over {nR}}
} 
We see from eq.\levelmatch,that in static gauge, the level matching condition for a single 
wound D-string is that $M - {\bar M}$ is an integer. For an n times wound string this 
difference should be divisible by n. 

Let us first look at the $Spin(32)/ Z_2$ Heterotic string in the sector $N=1$. The matrix 
model for this one is the D-string wound on a circle with no SO(32) Wilson line.
There are two sectors to consider corresponding to the O(1) holonomy around the circle. 
In one sector the fermions $\chi$ are periodic, in the other they are antiperiodic. 
Let us calculate the zero point energies. The rightmovers are easy. Here
 the bosons and fermions contribute equally but opposite so the zero point 
energy is zero. This means that $M$ is a non negative integer.
On the non-supersymmetric leftmoving side there 
are 8 bosonic fields $X^i$. There are also 32 fermions, $\chi$. 
We remember that the zero point energy is $-1 \over 24$ for a periodic 
boson, $1 \over 24$ for a periodic fermion and $-1 \over 48$ for an 
antiperiodic fermion.
Let us first calculate the zero point energy in the periodic sector.
It is easily seen to be 1. This means that $\bar M$ is an integer, which is 
at least 1. We see that from this sector we do not get any massless states.
In the antiperiodic sector we get a zero point energy of -1. This means 
that $\bar M$ is at least -1. Furthermore it follows from eq.\levelmatch\ 
that $\bar M$ is an integer. In other words we have to excite an even number 
of $\chi$ oscillators. This also follows from O(1) gauge invariance. 
In the antiperiodic sector we have massless states. We can excite either one X 
oscillator or two $\chi$ oscillators. These states are the usual states 
of Heterotic SO(32) string theory, the gravitons and gluons. Furthermore 
there is also a state with ${\bar M} = -1$. This state has negative energy
 according to eq.\energi. It is easy to dualize back to the original 
SO(32) Heterotic string theory to see what it corresponds to. It is a wound 
string on the lightlike circle. 

Let us now go to $N > 1$
As discussed in \Motlscrew, \BS\ and \DVV\ the matrices can configure 
themselves into long strings. Suppose we have a long D-string with winding 
number N. Level matching now requires the difference $M - {\bar M}$ to be 
divisible by N. States with $M - {\bar M} = 0$ have energies that go as 
$1 \over N$ as can be seen from eq.\energi. States with other values of 
$M - {\bar M}$ have energies of order one with respect to N. This is good, 
if we hope to recover the full ten dimensional theory for $N \rightarrow 
\infty$, since this shows that wound strings on the lightlike circle 
decouples in the large N limit. The negative energy state is also 
abandoned since -1 is not divisible by N. So for long strings the low energy 
states correspond to particle states in ten dimensional SO(32) Heterotic
string theory and there are no negative energy states.
Similar remarks apply to the $\e8$ case.

Another aspect to discuss is the GSO projection. For the SO(32) case 
there is no problem. The GSO projection for N=1 is simply given by 
the element 
$-1 \in O(1)$. This element multiplies all $\chi$ by -1 and leaves the 
other fields invariant. This is exactly how the GSO projection works 
for the SO(32) Heterotic string. For larger N the GSO projection 
 is similarly imposed by gauge invariance.

 For the Heterotic $ \e8 $ string there are two GSO projections. One for 
each set of 16 $\chi$. The element $-1 \in O(1)$ only takes care of one 
of these. It multiplies all 32 $\chi$ by -1. We still need another one. 
As noted in \Lubos\ this would be solved by level matching for odd $\tilde N$. 
This is 
because the rightmovers always have an integer number of excitations. 
Therefore there must be an even number of excitations of the antiperiodic 
leftmoving fermions.
 For a long string composed of an even number $\tilde N$ of D-strings we need 
a GSO projection that distinguishes the two sets of fermions. They are both 
periodic on going all the way around the long string. However a physical state 
still has to be invariant under worldsheet translations around the circle once. 
One set of fermions are antiperiodic under this translation. There has to be an 
even number of excitations of these. This is the origin of the 
GSO projection for even $\tilde N$. A discussion of this point can be found in 
\Banksrev.
 If there had been no SO(32) Wilson line 
the level matching condition eq.\levelmatch\ would have been enough to make the 
state invariant under one translation around the circle, because the $\tilde N$ 
units of momentum around the long string are evenly distributed with one unit 
per single circle. With the SO(32) Wilson line this is not the case 
automatically and we need the extra condition that an even number of the 
sets of fermions, which are antiperiodic on going once around the circle,
are excited. Together with the gauge invariance mentioned above this 
implies the full GSO condition.  

Let us now discuss how we can see from the matrix model that the $\e8$ 
theory develops an 11'th dimension at strong coupling, whereas the 
SO(32) theory does not. Both matrix models have 8 fields X, which 
correspond to the transverse dimensions. Then there is time and the 
lightlike circle. For the SO(32) case there should be no more dimensions.
 For the $\e8$ we expect an interval whose length grows with the string 
coupling, $\lam$. It is clear that this extra dimension comes from the 
O(N) gauge field, $A_{\mu}$. A Wilson line around the circle corresponds 
to position in the 11'th direction. With a O(N) Wilson line the charged fields 
in the 1+1 dimensional theory change boundary conditions around the 
circle. This will change the frequency of the associated oscillators 
and hence the zero point energy. This zero point energy acts as an 
induced potential for the Wilson line. The difference between the 
SO(32) and $\e8$ case is that for the former the potential locks the 
Wilson line at a special value, but for the latter the potential is flat.

Let us calculate some zero point energies to verify this picture.
 The rightmovers are easy. Here
 the bosons and fermions contribute equally but opposite so the zero point 
energy is zero.  We can apply oscillator raising operators to increase 
the energy. The increase depends on the periodicity of the oscillator
which is determined by the Wilson line. On the non-supersymmetric leftmoving 
side there 
are 8 bosonic fields $X^i$ in the symmetric part of $N \otimes N$ and 8 
fermions $\lam_{-}$ in the adjoint of O(N). They cancel each other 
except for 8N bosonic fields. There are also 32 fermions, $\chi$, in the 
N of O(N). We remember that the zero point energy (in units of 
$1 \over {R_s^{''}} $ for a wound string) is $-1 \over 24$ for a periodic 
boson, $1 \over 24$ for a periodic fermion and $-1 \over 48$ for an 
antiperiodic fermion.

Let us first calculate the zero point energy with no Wilson line, 
$A_{\mu}=0$. For the $\e8$ case there are 8N periodic bosons, 
16N periodic fermions and 16N antiperiodic fermions giving a total 
zero point energy of zero. For the SO(32) case we have 8N periodic 
bosons and 32N periodic fermions giving a total zero point energy of 
N.

Let us now put the Wilson line $-1 \in O(N)$ around the circle. This 
changes the periodicity of the fermions. For the $\e8$ case we still get 
zero. For the SO(32) case we now get -N. A more generic Wilson line 
will just give a result in between these two extreme Wilson lines. This 
is exactly what we expected. For the $\e8$ case it is always zero. For 
the SO(32) case it is smallest when the Wilson line is locked at -1. 
This state has negative energy and is the string wound on the 
lightlike circle as discussed above. 
It disappears when working with long strings instead.
To get massless states we have to raise some leftmoving 
oscillators. We can apply one X oscillator or two 
$\chi$ oscillators. The last states are the gluons in the adjoint 
of SO(32).

There is a more pictorial way of understanding all this, namely the 
T-dual version. Here the Wilson lines turn into positions of 0-branes 
on the interval in type IA. This system has been studied by 
\refs{\Kleb,\Ulf}. It was shown, by considering zero point energies for 
instance, that 8-branes repel 0-branes. For the $\e8$ case the 8-branes 
are split evenly among the two orientifold planes so the 0-branes feel 
no force. In the SO(32) case all 8-branes are gathered in one end(the 
rightmost) of the interval. The 0-branes will then be repelled to the 
other end(leftmost end). The excited $\chi$ oscillators correspond to 
strings stretched from the 0-branes to the 8-branes in the other end.       
The excited X oscillators correspond to strings winding once on the 
interval, starting from a 0-brane and ending on a 0-brane.
It is clear from this picture that two $\chi$ oscillators have the same 
energy as one X oscillator.

%%% ------------------------- CUT HERE ---------------------------------%

% ===================================================================== %
% Section (5) Type I Matrix Model
% ===================================================================== %
\newsec{Type I Matrix Model}

Since type I and SO(32) Heterotic string theory are dual in ten 
dimensions we also have a matrix model for type I. Let us start 
with type I with string scale $m_I$ and coupling $\lam_{I}$ on a lightlike 
circle of radius R. This is dual to SO(32) Heterotic theory with 
radius R, coupling $\lam = {1 \over \lam_{I} }$ and string scale 
$M = m_{I} \lam_{I}^{- \half}$. The matrix model for type I is thus 
again the theory of N D-strings wound on a circle with no Wilson line. 
The parameters of the type I theory where the D-strings live are given by 
substitution in eq.\dualto. Like before
\eqn\typeen{\eqalign{
\lam^{''} &\rightarrow 0 \cr
m_s^{''} &\rightarrow \infty \cr
R_s^{''}& = {\lam_{I} \over {m_I^2 R}}
}}
The gauge coupling is given from eq.\gaugekobling,
\eqn\koblingen{
{1 \over g^2} = { 1 \over {m_I^4 R^2}}.
}
The matrix model is thus a 1+1 dimensional O(N) theory on a circle of 
radius $\lam_{I} \over {m_I^2 R}$. Especially momentum around the circle 
has energy ${m_I^2 R}\over {\lam_{I}}$. This means, naively at least, that 
perturbative type I theory is reproduced by the dimensional reduction of 
this model to 0+1 dimensions. This is a quantum mechanics system!

In the T-dual version the restriction to quantum mechanics means that we
do not consider strings that go from one end of the interval to the other 
except for those needed to fulfill level matching for a long string. 
We remember there 
were two ways of doing this, either exciting one $X^i$ oscillator or two 
$\chi$ oscillators. In the type IA picture exciting an $X^i$ oscillator 
means exciting a string stretched around the interval coming back to the 
0-branes. The $\chi$ oscillator corresponds to a string stretched from 
a 0-brane to an 8-brane in the other end. We need two of these. It is
astonishing how much this resembles type I perturbation theory. The first 
kind of 0-brane corresponds to a closed string. The second kind to an 
open string with Chan-Paton factors. From this picture it is clear that 
a closed string can open up and become an open string.  
The fact that we need to excite 
oscillators mean that the theory is not quite the naive dimensional 
reduction. One has to keep track of other oscillators too. Hopefully 
it is still tractable.

Let us have a look at the sector N=1. Here we get the states $(8_v +
8_c) \otimes 8_v$ from the ``closed string sector'' and $8_v + 8_c$ 
in the adjoint of SO(32) from the open string sector (we discarded 
the negative energy state which is absent for a long string anyway). 
These are the 
massless states in type I. We see that we do not get any massive states 
in type I. This is not an immediate contradiction since massive states 
are unstable for finite coupling and are not asymptotic states.  
Exciting more oscillators we get states with energies that go like $1 \over 
{\lam_{I}}$. These correspond to D-strings of type I. It would be 
interesting to see from this picture of type I why fundamental strings can 
end on D-strings.

%------------------------------------------------------------------%
% section (6) Conclusions
%------------------------------------------------------------------%
\newsec{Conclusions}

We have seen how the two Heterotic strings have a matrix model 
description as a 1+1 dimensional O(N) gauge theory. This was derived 
for very special Wilson lines. It would be very interesting to 
understand the general case. It might be related to the recent 
discoveries by Connes, Douglas, Hull and Schwarz \refs{\Connes, \DougHull}.
We saw how the integer N of the matrix model was not just momentum around 
the lightlike circle but a generalized momentum eq.\skiftetmom. For 
general Wilson lines it is hard to get an integer out of a formula 
containing the Wilson line. It would be interesting to figure out 
what matrix models with non-integer N means, if they exist.

We also saw how the type I theory is described by a 1+1 dimensional 
O(N) gauge theory. For small coupling it reduces to a quantum 
mechanical problem. There is however a problem in that we have to 
keep an excited oscillator. This means a naive dimensional reduction 
to 0+1 dimensions is too simple. We have to keep a finite 
number of excited oscillators. It is still a quantum mechanical 
system however. 
 It would be very interesting 
to do a scattering calculation and compare to traditional type I 
results.

So far the only matrix model which is 0+1 dimensional is the 
original model of M-theory in the DLCQ \BFSS\ . One can do 
scattering calculations there but the problem is that there is 
only supergravity to compare with. It is not clear whether the 
system in the DLCQ is describable by supergravity and possible 
disagreements might be because we are outside the regime of 
validity of supergravity in this special kinematical situation.
In the type I model the situation is seemingly better in this 
respect since we have a string theory to compare with. 
Unfortunately it might not be better, since type I on a small 
circle is not perturbative \PolWit. So it might not be possible 
to compare with a string calculation. 
Certainly the arguments in this paper imply that the 1+1 
dimensional model describes type I in the DLCQ, where DLCQ 
is defined as a limit of an almost lightlike circle.
 For small $\lam_{I}$ 
the model becomes 0+1 dimensional. 
However defining the DLCQ as a limit of an 
almost lightlike circle the theory is Lorentz equivalent
 to type I on a very small circle. This is not perturbative 
for small $\lam_I$, so we have not derived agreement between 
perturbative type I and the 0+1 dimensional matrix model 
described here for finite N.  
 Following the philosophy 
of matrix theory so far we could hope that for large N all traces of the 
lightlike compactification disappears and type I with a small coupling would 
be perturbative. In this case the result would be that perturbative type I 
is reproduced by a large N quantum mechanics. We saw one hint of the 
disappearance of the lightlike circle for large N, namely the Heterotic string 
wound on the lightlike circle was not present for a long D-string in the Matrix model,
 even though it was present for a single D-string.

%%% ------------------------- CUT HERE ---------------------------------%
%=======================================================================%
% Acknowledgments
%=======================================================================%
\bigbreak\bigskip\bigskip
\centerline{\bf Acknowledgments}\nobreak
I wish to thank Ori Ganor for fruitful discussions and Jan Ambjorn 
for hospitality at the Niels Bohr Institute.
This work was supported by The Danish Research Academy.

%%% ------------------------- CUT HERE ---------------------------------%

\listrefs
\bye
\end